\documentclass[12pt,a4paper,notitlepage]{article}
\usepackage{amsmath}
\usepackage{epsfig}
\usepackage{graphicx}

\begin{document}

\author{Adam Bzdak and Micha\l\ Prasza\l owicz \\
{\footnotesize M. Smoluchowski Institute of Physics,}\\
{\footnotesize \ Jagellonian University, }\\
{\footnotesize \ ul. Reymonta 4, PL-30-059 Krak\'{o}w, Poland}}
\title{ \hfill{\small  TPJU-03/2003}\\
~~\\
An attempt to construct pion distribution amplitude \\
from the PCAC relation\\
in the nonlocal chiral quark model\\
~}
\maketitle

\begin{abstract}
Using the PCAC relation, we derive a compact formula for the pion decay
constant $F_{\pi}$ in the nonlocal chiral quark model. For practical
calculations this formula may be used both in the Minkowski and in the
Euclidean space. For the pion momentum $P_{\mu}\rightarrow0$ it reduces to
the well known expression derived earlier by other authors. Using a
generalized dipole Ansatz for the momentum dependence of the constituent
quark mass in the Minkowski space, we express $F_{\pi}^{2}$ in terms of a
single integral over the quark momentum fraction $u$. We interpret the
integrand as a pion distribution amplitude $\phi(u)$. We discuss its
properties and compare with the $\pi$DA's obtained in other models.
\end{abstract}

\section{Introduction\label{intro}}

Recent data from CLEO \cite{cleo} and E791\cite{e791} experiments triggered
a new wave of theoretical studies of the leading twist pion distribution
amplitude ($\pi $DA). On one side the data have been reanalyzed taking into
account NLO perturbative QCD effects, as well as nonperturbative effects
parameterized within the QCD light-cone sum rules \cite{SY},\cite{nico1}. On
the other hand nonperturbative models \cite{nico2}--\cite{mpar} and lattice
QCD \cite{lattice}--\cite{lattice3} have been employed to calculate the $\pi 
$DA from the relatively nonrestrictive physical assumptions. Here the dual
nature of the pion, being the quark--antiquark bound state and the Goldstone
boson of the broken chiral symmetry at the same time, makes such
calculations interesting by itself, even if the data is not yet decisive
enough to distinguish between different models.

Pion distribution amplitude is usually defined by means of the following
matrix element (see \emph{e.g.} \cite{Ball}): 
\begin{align}
\phi_{\pi}(u) & =\frac{1}{i\sqrt{2}F_{\pi}}\int\limits_{-\infty}^{\infty }%
\frac{d\tau}{\pi}e^{-i\tau(2u-1)(nP)}  \notag \\
& \left\langle 0\left| \bar{d}(n\tau)\rlap{/}n\gamma_{5}u\left(
-n\tau\right) \right| \pi^{+}(P)\right\rangle   \label{Fipidef}
\end{align}
in the light cone kinematics where two quarks separated by the light cone
distance $z=2\tau$ along the direction $n=(1,0,0,-1)$ are moving along the
light cone direction $\tilde{n}=(1,0,0,1)$ parallel to the total momentum $P$%
. Here $F_{\pi}=93$ MeV. In this kinematical frame any four vector $v$ can
be decomposed as: 
\begin{equation}
v^{\mu}=\frac{v^{+}}{2}\tilde{n}^{\mu}+\frac{v^{-}}{2}n^{\mu}+v_{\bot}^{\mu
}   \label{LC}
\end{equation}
with\quad$v^{+}=n\cdot v,\quad v^{-}=\tilde{n}\cdot v$, and the scalar
product of two four vectors reads: 
\begin{equation}
v\cdot w=\frac{1}{2}v^{+}w^{-}+\frac{1}{2}v^{-}w^{+}-\vec{v}_{\bot}\cdot 
\vec{w}_{\bot}.
\end{equation}
In Eq.(\ref{Fipidef}) the path ordered exponential of the gluon field,
required by the gauge invariance, has been omitted since we shall be working
in the effective quark model where the gluon fields have been integrated out.

In the local limit matrix element (\ref{Fipidef}) reduces to 
\begin{equation}
\left\langle 0\left| A_{\mu}^{a}(x)\right| \pi^{b}(P)\right\rangle
=-iP_{\mu}F_{\pi}\delta^{ab}e^{-iPx}   \label{Amuel}
\end{equation}
where 
\begin{equation}
A_{\mu}^{a}(x)=\bar{\psi}(x)\gamma_{\mu}\gamma^{5}\frac{\tau^{a}}{2}\psi(x) 
\label{Amu}
\end{equation}
is the properly normalized axial vector current.

In Refs.\cite{mpar} $\phi_{\pi}(u)$ has been calculated in the effective
chiral quark model in which quarks interact \emph{nonlocally} with an
external meson field 
\begin{equation}
U^{\gamma_{5}}(x)=e^{i\gamma_{5}\tau^{a}\pi^{a}(x)/F_{\pi}}   \label{Ug5}
\end{equation}
and acquire a momentum dependent constituent mass 
\begin{equation}
M(k)=M_{k}=MF(k)^{2}.   \label{Mk}
\end{equation}
$M$ is a constituent quark mass of the order of $350$ MeV and $F(k)$ is a
momentum dependent function such that $F(0)=1$ and $F(k^{2}\rightarrow
\infty)\rightarrow0$. Function $F(k)$ embodies nonperturbative effects due
to the nontrivial structure of the QCD\ vacuum. Indeed, $F(k)$ has been
explicitly derived within the instanton model \cite{inst1}. In Refs.\cite
{mpar} where the calculations were performed in the Minkowski space
(instanton model is inevitably formulated in the Euclidean metric) a
convenient Ansatz for $F(k)$ was used: 
\begin{equation}
F(k)=\left( \frac{-\Lambda^{2}}{k^{2}-\Lambda^{2}+i\epsilon}\right) ^{n}. 
\label{Fkdef}
\end{equation}
With this Ansatz $\phi_{\pi}(u)$ as well as higher twist $\pi$DA's were
calculated in Refs.\cite{mpar} and \cite{mpar1} respectively.

The problem is, however, that in the model with the nonlocal interaction
(and momentum dependent quark mass $M_{k}$) the axial current (\ref{Amu})
does not exhibit PCAC \cite{Holdom}--\cite{WBnl}. More drastically, a naive
vector current 
\begin{equation}
V_{\mu}^{a}(x)=\bar{\psi}(x)\gamma_{\mu}\frac{\tau^{a}}{2}\psi(x) 
\label{Vmu}
\end{equation}
is not conserved. In order to restore these properties extra currents have
to be added to $A_{\mu}^{a}$ and $V_{\mu}^{a}$ \cite{RiBall},\cite{Birse}.
These new pieces modify both model expressions for $F_{\pi}$ and for $%
\phi_{\pi}(u)$. While the formula for $F_{\pi}$ is well known in terms of
the Euclidean integral \cite{Birse},\cite{DiakPetFpi}: 
\begin{equation}
F_{\pi}^{2}=4N_{c}\int\frac{d^{4}k_{E}}{(2\pi)^{4}}\frac{%
M_{k}^{2}-k_{E}^{2}M_{k}M_{k}^{\prime}+k_{E}^{4}M_{k}^{\prime2}}{%
(k_{E}^{2}+M_{k}^{2})^{2}}   \label{Euclid}
\end{equation}
(here $M_{k}^{\prime}=dM_{k}/dk^{2})$ the form of the wave function has been
a subject of different studies with, however, contradictory results. For
example the distribution amplitude obtained in Ref.\cite{dor} is very close
to the asymptotic form 
\begin{equation}
\phi_{\pi}^{as}(u)=6u(1-u)   \label{as}
\end{equation}
where $u=k^{+}/P^{+}$ is the momentum fraction carried by the quark, whereas
in Refs.\cite{erad}--\cite{erab} $\phi_{\pi}(u)=1$.

In the present work we derive the Minkowski space formula for $F_{\pi}^{2}$
for the modified axial current replacing the naive current in Eq.(\ref{Amuel}%
). Our formula, when continued to the Euclidean space, reduces to Eq.(\ref
{Euclid}). However, when evaluated in the Minkowski space by methods
developed in Refs.\cite{mpar}, it can be represented as an integral over $du$
from an integrand which we interpret as $\phi_{\pi}(u)$. This function does
not resemble (\ref{as}) and is compatible rather with the constant wave
function of Refs.\cite{erad}--\cite{erab} than with the result obtained in
the same model \cite{mpar}, however, with the naive current (\ref{Amu}).

There are several comments which are due at this point. First of all it is
not clear how the modified current can be generalized to the bilocal
operator entering formula (\ref{Fipidef}). That is why it was argued in
Refs. \cite{dor}--\cite{erab} that rather than considering matrix elements
of the form (\ref{Fipidef}) or (\ref{Amuel}), one should calculate the whole
physical process in the effective model, impose Bjorken limit to make
contact with the expressions known from QCD and extract the distribution
amplitude. One has to note however, that the effective models are not valid
at large momenta which are needed to impose Bjorken limit. Moreover it is
not clear whether the distribution amplitudes defined that way are
universal. Secondly, arguments may be given that it is not necessary to
insist that the bilocals defining the distribution amplitudes must reduce to
the proper currents in the local limit\footnote{%
By local limit we understand the limt in which the fields in Eq.(1) are
taken in the same point $x$. There are still corrections due to the momentum
dependent constituent mass and nonlocal interactions.}. Indeed, as we shall
show below the naive bilocal (\ref{Fipidef}) reproduces the Pagels-Stokar
formula \cite{PaSt} for $F_{\pi}^{2}$: 
\begin{equation}
F_{\pi}^{2}=4N_{c}\int\frac{d^{4}k_{E}}{(2\pi)^{4}}\frac{M_{k}^{2}-\frac{1}{2%
}k_{E}^{2}M_{k}M_{k}^{\prime}}{(k_{E}^{2}+M_{k}^{2})^{2}}   \label{PagSto}
\end{equation}
which was obtained from the Ward-Takahshi identities.

\section{Currents in the nonlocal models\label{currents}}

Let us consider the model defined by an action \cite{bochum},\cite{mpar}: 
\begin{equation*}
S=\,\int (dk)\bar{\psi}(k)\left( \rlap{/}{k}-m\right) \psi (k)-M\int (dk\,dl)%
\bar{\psi}(l)F(l)U^{\gamma _{5}}(l-k)F(k)\psi (k).
\end{equation*}
Here, following \cite{WBnl} $(dk)=d^{4}k/(2\pi )^{4}$ \emph{etc.}, and $%
(dx)=d^{4}x$. Equations of motion for the quark fields read 
\begin{align}
\rlap{/}{k}\psi (k)& =M\int (dl)F(k)U^{\gamma _{5}}(k-l)F(l)\psi (l)+m\psi (k), 
\notag \\
\bar{\psi}(k)\rlap{/}{k}& =M\int (dl)\bar{\psi}(l)F(l)U^{\gamma _{5}}(l-k)F(k)+m%
\bar{\psi}(k).  \label{EoM}
\end{align}
To get the equation of motion for the $U^{\gamma _{5}}$ field let us expand (%
\ref{Ug5}) 
\begin{equation}
U^{\gamma _{5}}(l-k)=(2\pi )^{4}\delta ^{(4)}(l-k)+\frac{i}{F_{\pi }}\gamma
^{5}\tau ^{c}\pi ^{c}(l-k)+\ldots   \label{Uexpand}
\end{equation}
and the equation of motion gives a constraint 
\begin{equation}
\int (dl)\,\bar{\psi}(l+k)F(l+k)\gamma ^{5}\tau ^{a}F(l)\psi (l)=0.
\label{constr}
\end{equation}

It is easy to verify that the naive vector current (\ref{Vmu}) is not
conserved \cite{Birse},\cite{WBnl}. In order to restore current
conservation, the following two currents have to be added to $V_{\mu }^{a}$
(in momentum space) 
\begin{equation}
\tilde{V}_{\mu }^{a}(P)=V_{\mu }^{a}(P)+R_{\mu }^{a}(P)+L_{\mu }^{a}(P)
\end{equation}
with left and right currents defined as 
\begin{align}
L_{\mu }^{a}(P)& =iM\int (dx\,dy\,dz)\,\int\limits_{x}^{z}ds_{\mu }e^{iPs}\,%
\bar{\psi}(x)F(x-z)T^{a}U^{\gamma _{5}}(z)F(z-y)\psi (y),  \notag \\
R_{\mu }^{a}(P)& =iM\int (dx\,dy\,dz)\,\int\limits_{z}^{y}ds_{\mu }e^{iPs}\,%
\bar{\psi}(x)F(x-z)U^{\gamma _{5}}(z)T^{a}F(z-y)\psi (y)  \label{LR}
\end{align}
where $T^{a}=\tau ^{a}/2$. Accordingly the modified axial current reads: 
\begin{equation}
\tilde{A}_{\mu }^{a}(P)=A_{\mu }^{a}(P)+R_{\mu }^{a}(P)-L_{\mu }^{a}(P)
\label{Atilde}
\end{equation}
with $T^{a}=\gamma _{5}\tau ^{a}/2$. The integral $ds_{\mu }$ should be
understood as an integral over the path connecting points $z$ and $y$ or $x$%
. This prescription makes the $L_{\mu }^{a}$ and $R_{\mu }^{a}$ currents
path dependent \cite{Birse} (strictly speaking the transverse part is not
fixed).

The divergence of the modified vector current is, however, path independent
and takes the following form: 
\begin{equation}
P^{\mu}\tilde{V}_{\mu}^{a}(P)=M\int(dk\,dl)\,\bar{\psi}(k)F(k)\left[ \frac{%
\tau^{a}}{2},U^{\gamma_{5}}(k-l+P)\right] F(l)\psi(l).   \label{PmuVt}
\end{equation}
This is immediately zero for the baryon current ($\tau^{a}=1$). For the
isospin current we can expand $U^{\gamma_{5}}$ (\ref{Uexpand}) 
\begin{equation}
\left[ \frac{\tau^{a}}{2},U^{\gamma_{5}}(k-l+P)\right] =\frac{1}{F_{\pi}}%
\gamma^{5}\tau^{a}\epsilon^{abc}\pi^{c}(k-l+P)+\ldots
\end{equation}
and (\ref{PmuVt}) vanishes due to the constraint (\ref{constr}). For the
axial current we get 
\begin{align}
P^{\mu}\tilde{A}_{\mu}^{a}(P) & =-m\int(dk)\bar{\psi}(k)\gamma_{5}\tau
^{a}\psi(k+P)  \notag \\
& -M\int(dk\,dl)\bar{\psi}(k)F(k)\left\{ \gamma_{5}\frac{\tau^{a}}{2}%
,U^{\gamma_{5}}(k-l+P)\right\} _{+}F(l)\psi(l).   \label{dAtilde}
\end{align}
By expanding $U^{\gamma_{5}}$ (\ref{Uexpand}) we arrive at 
\begin{align}
P^{\mu}\tilde{A}_{\mu}^{a}(P) & =-m\int(dk)\bar{\psi}(k)\gamma_{5}\tau
^{a}\psi(k+P)  \notag \\
& -M\int(dk)\bar{\psi}(k)F(k)\gamma_{5}\tau^{a}F(k+P)\psi(k+P)  \notag \\
& -i\frac{M}{F_{\pi}}\int(dk\,dl)\bar{\psi}(k)F(k)F(l)\psi(l)\,\pi
^{a}(k-l+P)+\ldots   \label{dAtilde1}
\end{align}
which is the proper PCAC formula (note that the second term vanishes due to (%
\ref{constr})).

In order to calculate $F_{\pi}$ we can either use Eq.(\ref{Amuel}) with $%
A_{\mu}^{a}\rightarrow\tilde{A}_{\mu}^{a}$ \cite{Bzdak} or use the PCAC
relation 
\begin{equation}
\left\langle 0\left| i\partial^{\mu}\tilde{A}_{\mu}^{a}(x)\right| \pi
^{b}(P)\right\rangle =-iP^{2}F_{\pi}\delta^{ab}e^{-iPx}   \label{PCAC}
\end{equation}
which is what we are going to do in this work. Notice, that we have to
calculate the matrix element in Eq.(\ref{PCAC}) off-shell, extract the
leading power in $P^{2}$ and take the limit $P^{2}\rightarrow0$.

\section{Decay constant and the distribution amplitude}

\subsection{Matrix elements}

%
\vspace{0.5cm} 
\begin{figure}[h]
\begin{center}
\includegraphics[scale=0.7]{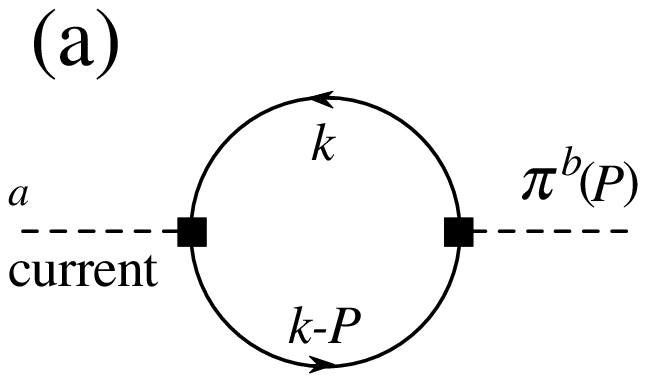} \hspace{0.9cm} 
\includegraphics
[scale=0.7]{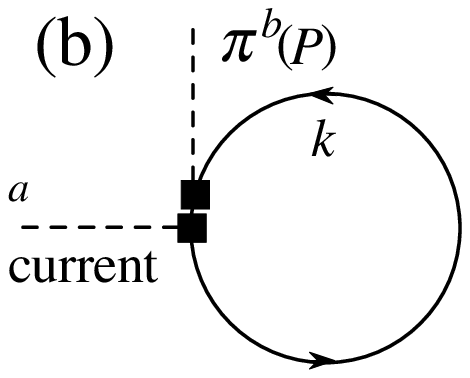} \hspace{0.9cm} %
\includegraphics[scale=0.7]{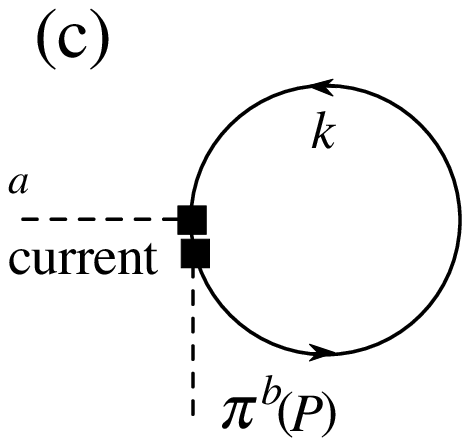}
\end{center}
\caption{{\protect\footnotesize Diagrams contributing to the matrix element
of Eq.{\ (21)}. Black squares denote $\protect\tau ^{a,b}\protect\gamma _{5}$%
. }}
\label{diags}
\end{figure}
%
%

There are three contributions to the matrix element of Eq.(\ref{dAtilde})
depicted in Fig.\ref{diags}: one from the first term of expansion (\ref
{Uexpand}) and two (which by the anticommutation rule reduce to one term,
see Eq.(\ref{dAtilde1})) from the term in (\ref{Uexpand}) involving one pion
field. Adding all of them we get 
\begin{align}
\left\langle 0\left| i\partial^{\mu}\tilde{A}_{\mu}^{a}(z)\right| \pi
^{b}(P)\right\rangle & =-\frac{8N_{c}}{F_{\pi}}\delta^{ab}e^{-iPz}\times \\
& \int(dk)\left[ M_{k}M_{k-P}\frac{k(P-k)+M_{k}M_{k-P}}{%
(k^{2}-M_{k}^{2})((k-P)^{2}-M_{k-P}^{2})}+\frac{M_{k}^{2}}{k^{2}-M_{k}^{2}}%
\right] .  \notag
\end{align}
Symmetrizing the last term with respect to the change of variables $%
k\rightarrow k-P$, adding all terms and comparing with Eq.(\ref{PCAC}) we
arrive at 
\begin{equation}
F_{\pi}^{2}=-i4N_{c}\frac{1}{P^{2}}\int(dk)\frac{\left[ M_{k}(k-P)_{\mu
}-M_{k-P}k_{\mu}\right] ^{2}}{(k^{2}-M_{k}^{2})((k-P)^{2}-M_{k-P}^{2})}. 
\label{FullFpi}
\end{equation}
By expanding Eq.(\ref{FullFpi}) in powers of $P^{2}$ we recover the
Minkowski version of Eq.(\ref{Euclid}). Indeed, by changing the variables: $%
k\rightarrow k+P/2$ we get 
\begin{equation}
F_{\pi}^{2}=-i4N_{c}\frac{1}{P^{2}}\int(dk)\frac{1}{2}\frac{\left[
M_{k+P/2}(k-\frac{P}{2})_{\mu}-M_{k-P/2}(k+\frac{P}{2})_{\mu}\right] ^{2}}{%
((k+\frac{P}{2})^{2}-M_{k+P/2}^{2})((k-\frac{P}{2})^{2}-M_{k-P/2}^{2})}. 
\label{FullFpi0}
\end{equation}
In fact an expression identical to Eq.(\ref{FullFpi0}) appears in the axial
and pseudoscalar corellators derived within the instanton model of the QCD
vacuum \cite{DiakPetFpi}.

Noting that 
\begin{equation}
M_{k\pm P/2}=M_{k}\pm(kP)M_{k}^{\prime}+\frac{P^{2}}{4}M_{k}^{\prime}+\ldots
\end{equation}
where $^{\prime}$ denotes $d/dk^{2}$ we have 
\begin{equation}
F_{\pi}^{2}=-i4N_{c}\frac{1}{P^{2}}\int(dk)\frac{%
P^{2}M_{k}^{2}-4(Pk)^{2}M_{k}M_{k}^{\prime}+4k^{2}(kP)^{2}M_{k}^{\prime2}}{%
((k+\frac{P}{2})^{2}-M_{k+P/2}^{2})((k-\frac{P}{2})^{2}-M_{k-P/2}^{2})}. 
\label{fpi6}
\end{equation}
Since under the integral $k_{\mu}k_{\nu}\rightarrow\frac{1}{4}g_{\mu\nu}k^{2}
$ (plus a term proportional to $P_{\mu}P_{\nu}$ which we may safely neglect)
equation (\ref{fpi6}) transforms into 
\begin{align}
F_{\pi}^{2} & =-i4N_{c}\int(dk)\frac{M_{k}^{2}-k^{2}M_{k}M_{k}^{%
\prime}+k^{4}M_{k}^{\prime2}}{((k+\frac{P}{2})^{2}-M_{k+P/2}^{2})((k-\frac{P%
}{2})^{2}-M_{k-P/2}^{2})}  \notag \\
& =-i4N_{c}\int(dk)\frac{M_{k}^{2}-k^{2}M_{k}M_{k}^{\prime}+k^{4}M_{k}^{%
\prime2}}{(k^{2}-M_{k}^{2})^{2}}.   \label{fullFpi1}
\end{align}

On the other hand, matrix element of the naive current (\ref{Amu}), gives 
\cite{mpar} 
\begin{equation}
F_{\pi}^{2}P_{\mu}=-i4N_{c}\int(dk)\sqrt{M_{k}M_{k-P}}\frac{M_{k-P}k_{\mu
}+M_{k}(P_{\mu}-k_{\mu})}{\left( k^{2}-M_{k}^{2}\right) \left(
(k-P)^{2}-M_{k-P}^{2}\right) }   \label{PagStoMink}
\end{equation}
which by the same steps which led from Eq.(\ref{FullFpi}) to (\ref{fullFpi1}%
) reduces equation (\ref{PagStoMink}) to the Minkowski version of the
Pagels-Stokar formula (\ref{PagSto}).

\subsection{Calculation of the loop integral}

In order to calculate the loop integral in Eq.(\ref{FullFpi}) with $M_{k}$
given by Eqs.(\ref{Mk},\ref{Fkdef}) we shall introduce the light-cone
parameterization of the momenta (\ref{LC}) with 
\begin{equation}
d^{4}k=\frac{P^{+}}{2}du\,dk^{-}d^{2}\vec{k}_{\bot}
\end{equation}
where $k^{+}=uP^{+}$. The method of evaluating $dk^{-}$ integral, taking the
full care of the momentum mass dependence, has been given in \cite{mpar}. To
evaluate $dk^{-}$ integral we have to find the poles in the complex $k^{-}$
plane. It is important to note that the poles come only from the momentum
dependence in the denominators of Eqs.(\ref{FullFpi},\ref{PagStoMink}). This
means that the position of the poles is given by the zeros of denominator,
that is by the solutions of the equation 
\begin{equation}
k^{2}-M^{2}\left( \frac{\Lambda^{2}}{k^{2}-\Lambda^{2}+i\epsilon}\right)
^{4n}+i\epsilon=0.   \label{EqGz}
\end{equation}
This equation is equivalent to 
\begin{equation}
G(z)=z^{4n+1}+z^{4n}-r^{2}=\prod_{i=1}^{4n+1}(z-z_{i}),   \label{G}
\end{equation}
with $z=k^{2}/\Lambda^{2}-1+i\epsilon$ and $r^{2}=M^{2}/\Lambda^{2}$. For $%
r^{2}\neq0$ (or finite $\Lambda$) equation (\ref{G}) has $4n+1$
nondegenerate solutions which we denote $z_{i}$. Equation (\ref{EqGz})
should be understood as an equation for $k_{i}^{-}=k^{-}(z_{i})$. In general
case $4n$ of $z_{i}$'s can be complex and the care must be taken about the
integration contour in the complex $k^{-}$ plane. Because of the imaginary
part of the $z_{i}$'s, the poles in the complex $k^{-}$ plane can drift
across Re$k^{-}$ axis. In this case the contour has to be modified in such a
way that the poles are not allowed to cross it. This follows from the
analyticity of the integrals in the $\Lambda$ parameter and ensures the
vanishing of $\pi$DA's in the kinematically forbidden regions. The results
are expressed as sums over $z_{i}$'s which have to be found numerically.

In order to avoid spurious divergences coming from $k^{-}$ in the numerator
of Eq.(\ref{fullFpi1}) we shall make use of the Lorentz invariance, writing 
\begin{equation}
F_{\pi}^{2}=-i4N_{c}\frac{1}{P^{2}}I_{\mu\nu}g^{\mu\nu}.
\end{equation}
Since $I_{\mu\nu}$ vanishes for $P_{\mu}\rightarrow0$ (see (\ref{FullFpi})
and (\ref{fullFpi1})) we have that 
\begin{equation}
I_{\mu\nu}=A(P^{2})P_{\mu}P_{\nu}+\frac{1}{4}B(P^{2})P^{2}g_{\mu\nu}
\end{equation}
with 
\begin{equation}
A(P^{2})\rightarrow A,\quad B(P^{2})\rightarrow B\quad\text{for}\quad
P^{2}\rightarrow0.
\end{equation}
Then 
\begin{equation}
F_{\pi}^{2}=-i4N_{c}(A+B)   \label{sumA4B}
\end{equation}
Hence we have to calculate 2 integrals: 
\begin{equation}
A=\frac{1}{P^{+\,2}}n^{\mu}n^{\mu}I_{\mu\nu},\;B=-\frac{4}{P^{2}}%
\varepsilon_{\bot}^{\mu}\varepsilon_{\bot}^{\nu}I_{\mu\nu}. 
\label{integrals}
\end{equation}
The result reads 
\begin{align}  \label{AB}
A & =-\frac{i}{16\pi^{2}}M^{2}\int\limits_{0}^{1}du%
\sum_{i,k=1}^{4n+1}f_{i}f_{k}(z_{k}^{2n}\bar{u}+z_{i}^{2n}u)^{2}\times\ln%
\left( 1+z_{i}\bar {u}+z_{k}u\right) , \\
B & =-\frac{i}{16\pi^{2}}\frac{2M^{2}\Lambda^{2}}{P^{2}}\int%
\limits_{0}^{1}du\sum_{i,k=1}^{4n+1}f_{i}f_{k}(z_{k}^{2n}-z_{i}^{2n})^{2}%
\left( (1+z_{i}\bar{u}+z_{k}u)-\frac{P^{2}}{\Lambda^{2}}u\bar{u}\right) 
\notag \\
& \hspace{6cm}\times\ln\left( (1+z_{i}\bar{u}+z_{k}u)-\frac{P^{2}}{%
\Lambda^{2}}u\bar{u}\right) .  \notag
\end{align}
Here $\bar{u}=1-u$ and 
\begin{equation}
f_{i}=\prod \limits _{\substack{ k=1  \\ k\neq i}}^{4n+1}\frac{1}{%
z_{i}-z_{k}}=\prod \limits _{\substack{ k=1  \\ k\neq i}}^{4n+1}\frac{1}{%
z_{k}-z_{i}}
\end{equation}
for which the following identities hold \cite{mpar} 
\begin{equation}
\sum\limits_{i=1}^{4n+1}f_{i}z_{i}^{m}=\left\{ 
\begin{array}{ccc}
0 &  & m<4n \\ 
&  &  \\ 
1 &  & m=4n
\end{array}
\right. .
\end{equation}
As seen from Eq.(\ref{AB}) the first term in $B$ is singular as $%
P^{2}\rightarrow0$ in apparent contradiction with the finiteness of $%
F_{\pi}^{2}$. However, the function 
\begin{equation}
\phi_{\inf}(u)=-\frac{N_{c}M^{2}}{2\pi^{2}}%
\sum_{i,k=1}^{4n+1}f_{i}f_{k}(z_{k}^{2n}-z_{i}^{2n})^{2}(1+z_{i}\bar{u}%
+z_{k}u)\ln(1+z_{i}\bar{u}+z_{k}u)
\end{equation}
vanishes when integrated over $du$. Hence the finite formula for $F_{\pi}^{2}
$ reads 
\begin{equation}
F_{\pi}^{2}=-\frac{N_{c}M^{2}}{4\pi^{2}}\int\limits_{0}^{1}du\sum
_{i,k=1}^{4n+1}f_{i}f_{k}\left[ (z_{k}^{2n}\bar{u}%
+z_{i}^{2n}u)^{2}-2(z_{k}^{2n}-z_{i}^{2n})^{2}u\bar{u}\right] \ln\left(
1+z_{i}\bar{u}+z_{k}u\right) .   \label{fullFpi2}
\end{equation}
This allows us to define the distribution amplitude 
\begin{equation}
\tilde{\phi}_{\pi}(u)=-\frac{N_{c}M^{2}}{4\pi^{2}F_{\pi}^{2}}\sum
_{i,k=1}^{4n+1}f_{i}f_{k}\left[ (z_{k}^{2n}\bar{u}%
+z_{i}^{2n}u)^{2}-2(z_{k}^{2n}-z_{i}^{2n})^{2}u\bar{u}\right] \ln\left(
1+z_{i}\bar{u}+z_{k}u\right) .   \label{Phitld}
\end{equation}
Let us recall that the distribution amplitude defined by means of the naive
axial current (\ref{Amu}) reads \cite{mpar} 
\begin{equation}
\phi_{\pi}(u)=-\frac{N_{c}M^{2}}{4\pi^{2}F_{\pi}^{2}}\sum%
\limits_{i,k}f_{i}f_{k}\,(z_{i}^{n}z_{k}^{3n}\bar{u}+z_{i}^{3n}z_{k}^{n}u)%
\ln\left( 1+z_{i}\bar{u}+z_{k}u\right) .   \label{oldPhi}
\end{equation}

\subsection{Numerical results}

%
\vspace{0.5cm} 
\begin{figure}[h]
\begin{center}
\includegraphics[scale=0.7]{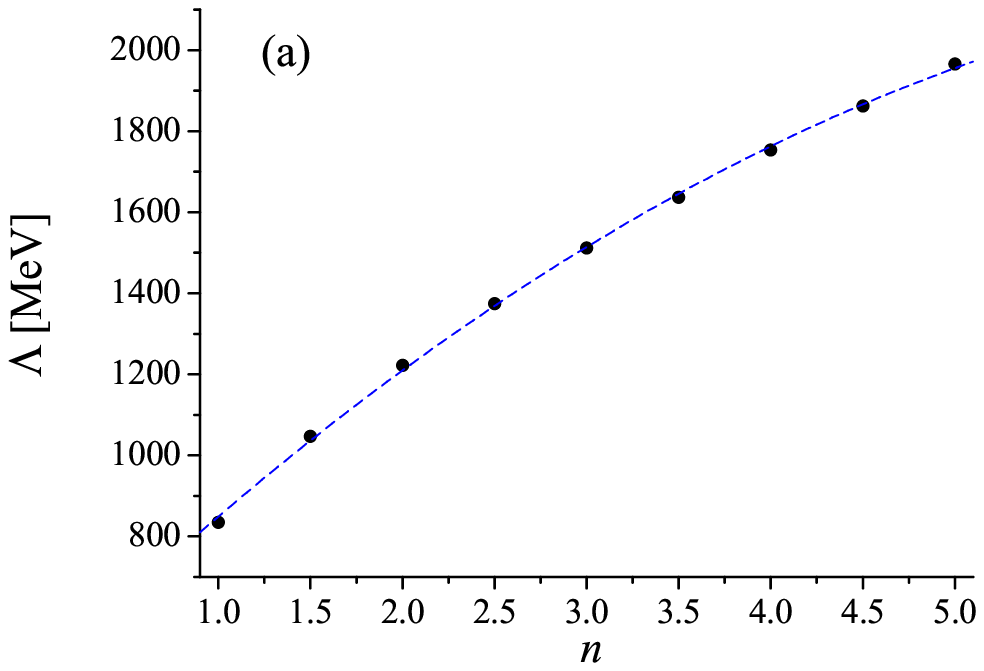} 
\includegraphics[scale=0.7]{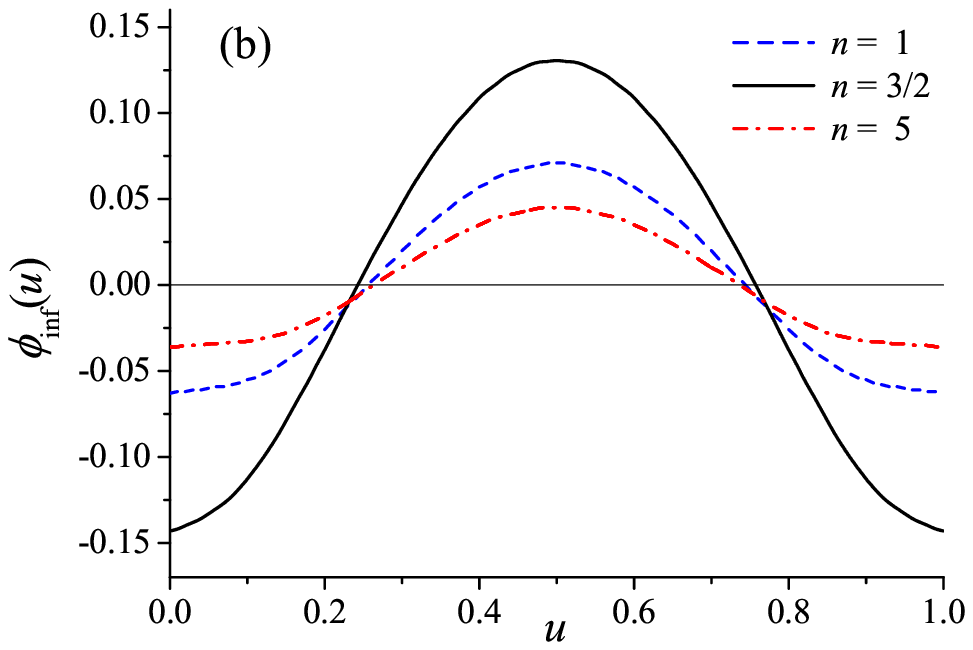}
\end{center}
\caption{{\protect\footnotesize {\ a) dots: cutoff parameter $\Lambda$ for
different $n$ (Eq.(8)) and for $M=350$~MeV, dahed line: fit described in the
text; b) function $\protect\phi_{\inf}(u)$ for $n=3/2,$ $3$ and $5$.}}}
\label{LG}
\end{figure}
%

Condition (\ref{Euclid}), or equivalently (\ref{fullFpi2}) provide a
relation between parameter $\Lambda$, constituent mass $M$ and power $n$
from Eq.(\ref{Fkdef}). Throughout this paper we shall use $M=350$ MeV. The
value of parameter $\Lambda=\Lambda(n)$ obtained from Eq.(\ref{fullFpi2}),
or from Eq.(\ref{Euclid}) after continuation of the cutoff formula (\ref
{Fkdef}) to the Euclidean metric, is depicted in Fig.\ref{LG}.a. It is
interesting to note, that our formula (\ref{fullFpi2}) for $F_{\pi}^{2}$,
unlike equation (\ref{PagStoMink}), does allow for half integer $n$'s. An
approximate relation, depicted by a dashed line in Fig.\ref{LG}.a holds 
\begin{equation*}
\Lambda\lbrack\text{MeV}]=432.82+444.61\,n-28.02\,n^{2}. 
\end{equation*}
The local current (\ref{Amu}) contributes, through Eq.(\ref{PagSto}),
approximately 70\% to the total normalization.

Having fixed $\Lambda$ for given $n$, we can calculate the distribution
amplitude as defined by Eq.(\ref{Phitld}). However, before doing this we
have to check whether the formally divergent part, given as an integral over 
$u$ from the function $\phi_{\inf}(u),$ vanishes. We have checked
numerically that this is indeed the case. Function $\phi_{\inf}(u)$ is
plotted in Fig.\ref{LG}.b for $n=3/2,$ $3$ and $5$.

\begin{figure}[h]
\begin{center}
\includegraphics[scale=0.7]{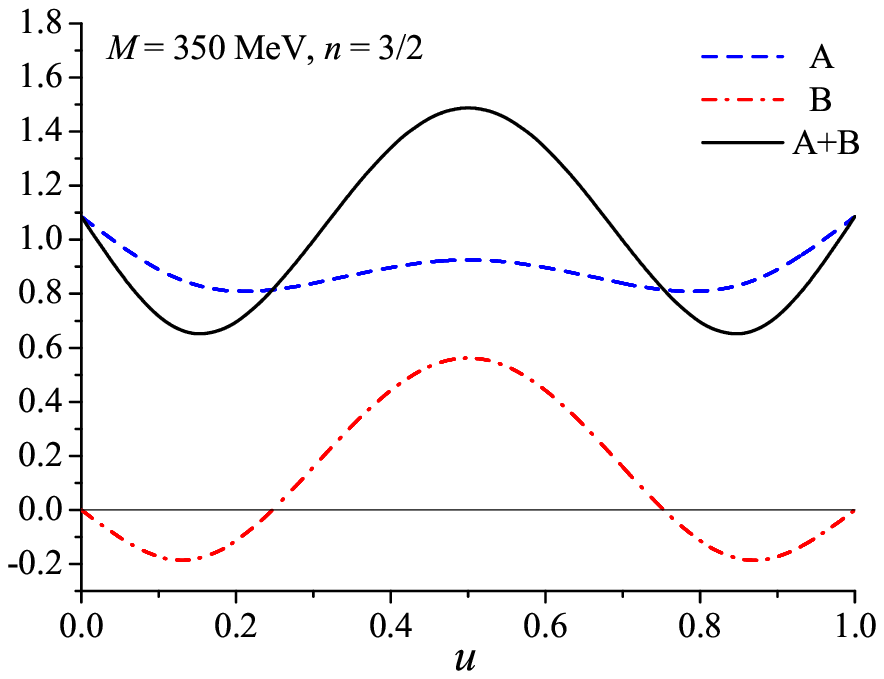} 
\includegraphics[scale=0.7]{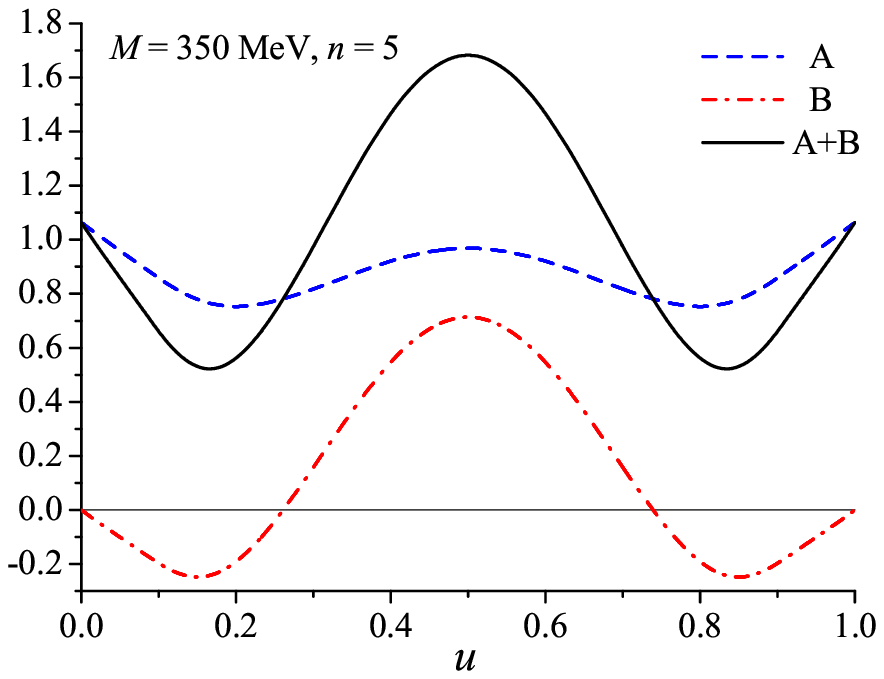}
\end{center}
\caption{{\protect\footnotesize {\ Function $\tilde{\protect\phi}_{\protect%
\pi}$ for $n=3/2$ and $n=5$ with two contributions $A$ and $B$ given by
Eqs.(39).}}}
\label{func}
\end{figure}

Next, in Fig.\ref{func} we plot the distribution amplitude $\tilde{\phi}%
_{\pi }(u)$ for $n=3/2$ and $n=5$ (solid lines) together with the
contributions from integrals $A$ and $B$ (\ref{sumA4B}). We see that the
contribution from $A$ is relatively flat and does not vanish at the end
points. The contribution from $B$ vanishes at the end points and is even
negative in their vicinity. There is not much difference between the two
cases $n=3/2$ and $n=5$, although one may say that the smaller $n$ the
flatter $\tilde{\phi}_{\pi}$.

%
\vspace{0.5cm} 
%

\begin{figure}[h]
\begin{center}
\includegraphics[scale=0.7]{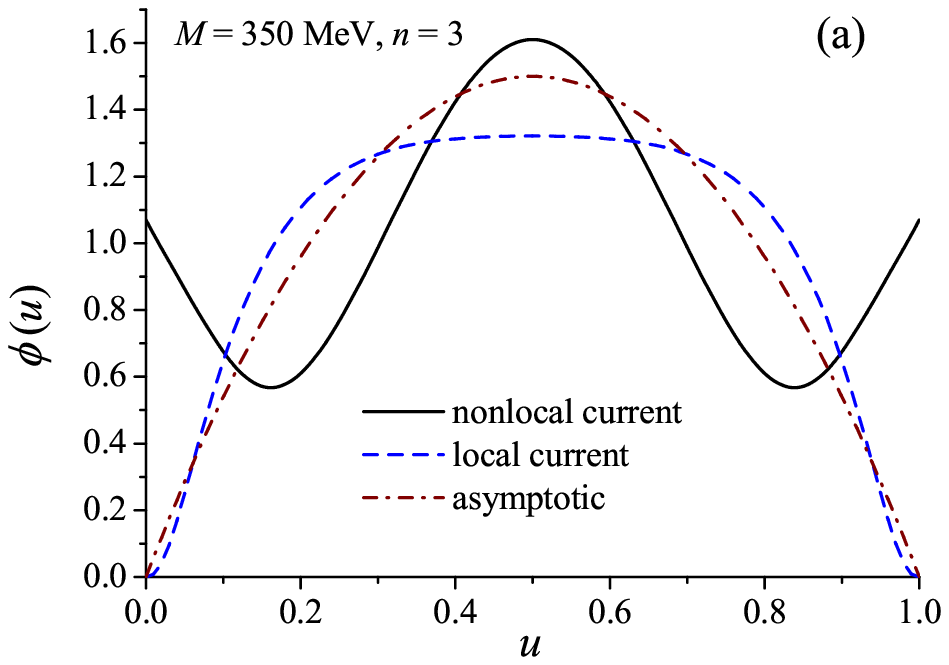} 
\includegraphics
[scale=0.7]{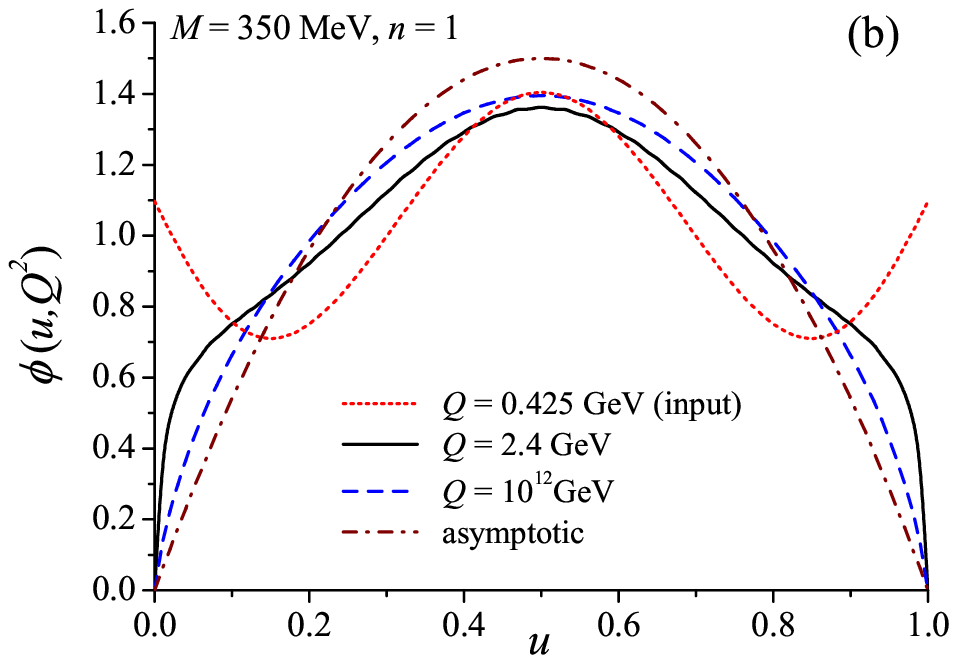}
\end{center}
\caption{{\protect\footnotesize {\ a): Functions $\tilde{\protect\phi}_{%
\protect\pi}$ (Eq.(44), solid), $\protect\phi_{\protect\pi}$ (Eq.(45),
dashed) for $n=3$ and asymptotic distribution amplitude (Eq.(11),
dash-dotted); b): Evolustion of $\tilde{\protect\phi}_{\protect\pi}$ from
the initial scale $Q=425$~MeV (dotted line) to the CLEO point $Q=2.4$~GeV
(solid) and for $Q=10^{12}$~GeV (dashed) for $n=1$.}}}
\label{comp}
\end{figure}

In Fig.\ref{comp}.a we plot for comparison function $\tilde{\phi}_{\pi }(u)$
(\ref{Phitld}), $\phi _{\pi }(u)$ (\ref{oldPhi}) corresponding to the naive
axial current (\ref{Amu}) for $M=350$ MeV and $n=3$, together with the
asymptotic distribution amplitude (\ref{as}). One should note that while
model distributions are defined as some low normalization scale $Q^{2}=\mu
^{2}$, $\phi _{\pi }^{as}(u)$ corresponds to the limit $Q^{2}\rightarrow
\infty $. Indeed, the leading twist distribution amplitude can be expanded
in terms of the Gegenbauer polynomials 
\begin{equation}
\phi _{\pi }(u;Q^{2})=6u(1-u)\left[ 1+\sum\limits_{n=2,4\ldots }^{\infty
}a_{n}(Q^{2})C_{n}^{3/2}(2u-1)\right]   \label{Gegen}
\end{equation}
where $a_{n}(Q^{2})\rightarrow 0$ in the large $Q^{2}$ limit \cite{ERBL}. It
is important no notice that $a_{n}(Q^{2})$ tend to zero monotonically, so
that they cannot change the sign.

%
\vspace{0.5cm} 
%

As soon as we switch on the QCD evolution, distribution amplitude $\tilde
{\phi}_{\pi}(u,Q^{2})$ changes the shape and it goes to zero at the end
points. This evolution is plotted in Fig.\ref{comp}.b for $n=1$, assuming 2
light flavors, $\Lambda_{QCD}=175$ MeV, and initial scale $\mu=425$ MeV.
This initial scale has been adjusted in such a way, that the second
Gegenbauer coefficient $a_{2}$, when evolved to the CLEO point $Q=2.4$ GeV,
gives $a_{2}(2.4)=0.15$ as indicated by the analysis of Ref.\cite{SY}. For
the normalization scale as large as $10^{12}$GeV the evolved distribution is
slowly approaching the asymptotic one.

\section{Summary and discussion}

In the present paper we have derived a compact formula (\ref{FullFpi}) for
the pion decay constant in the nonlocal chiral quark model. In order to
define $F_{\pi}$ we have used the full nonlocal axial current (\ref{Atilde})
and the PCAC relation (\ref{PCAC}). Equation (\ref{FullFpi}), when expanded
in the pion momentum $P_{\mu}$, reduces to the well known \cite{Birse},\cite
{DiakPetFpi} formula (\ref{fullFpi1}). The advantage of Eq.(\ref{FullFpi})
consists in the fact that it can be evaluated in the Minkowski space with a
suitable Ansatz for the momentum dependence of the constituent mass (\ref
{Fkdef}). By integrating (\ref{FullFpi}) over $dk^{-}$ and $d^{2}\vec{k}%
_{\bot}$ we are left with a $du$ integral over the function $\tilde{\phi}%
_{\pi}(u)$ which we \emph{interpret} as a pion distribution amplitude (\ref
{Phitld}).

As mentioned at the end of Sect.\ref{intro} it is not clear how to extend
the local current (\ref{Atilde}) to the bilocal operator like the one
entering formula (\ref{Fipidef}). Therefore our definition of the pion
distribution amplitude may be not correct. However, it is worth to note that
the shape of our distribution amplitude resembles a constant DA obtained by
a consistent use of the Ward-Takahashi identities \cite{erad}--\cite{erab},
rather than the DA calculated in the instanton model of the QCD vacuum in
Ref.\cite{dor}, although in both approaches full nonlocal currents have been
used.

Unfortunately, the $\pi$DA derived here and in Refs.\cite{erad}--\cite{erab}
is probably phenomenologically unacceptable. That is because the detailed
analysis of the CLEO data indicates that the coefficient $a_{4}(2.4$ GeV$)$
is \emph{negative} \cite{SY},\cite{nico1} and possibly as large as $%
a_{2}(2.4 $ GeV$)$ \cite{nico1}. In our case, however, $a_{4}$ is always 
\emph{positive}. The same concerns the constant $\pi$DA. In this respect $\pi
$DA derived in by the same methods in Refs.\cite{mpar} using the bilocal
operator (\ref{Fipidef}) with no extra pieces corresponding to the nonlocal
currents (\ref{LR}) fits the data much better. That is because, similarly to
the results of Refs.\cite{nico2}, it exhibits a shallow minimum around $%
u=1/2 $ which generates negative $a_{4}.$

As already mentioned above, there is a problem how to define the
distribution amplitudes in the effective models of QCD. This is due to the
fact that the QCD currents and the model currents are not the same. One way
would be to perform factorization and large $Q^{2}$ expansion in QCD and
then parameterize the nonperturbative matrix elements by a set of unknown
distribution amplitudes. To calculate these matrix elements an effective
model, like the one discussed here, is used. Considering operators as
obtained from QCD leads to the violation of PCAC and, in the worse case, to
the violation of the gauge invariance at the level of the effective model.
Another method consists in performing factorization and large $Q^{2}$
expansion directly in the effective model. This is possible, since the
degrees of freedom of the effective models discussed here are, at least as
the quantum numbers are concerned, identical to the degrees of freedom of
QCD (except for gluons, which are not present in the former case). This
means, however, that the low energy model has to be applied to the processes
with large momenum transfer. Since the currents of the effetive models are
not the same as in QCD, extra pieces contributing to the DA's, as compared
to the previous method, are present. Although in this work we have not
calculated the physical process and have not implemented the Bjorken limit,
our approach is in our opinion equivalent, since we have considered the
matrix element (\ref{Amuel}) of the full current (\ref{Atilde}). Our results
indicate that these two methods lead to completely different DA's . The
first method gives the $\pi$DA resembling the asymptotic distribution,
whereas the second approach generates the DA which is compatible with a
constant.

\bigskip

\begin{quote}
\textbf{Acknowledgments}
\end{quote}

It is a pleasure to dedicate this work to Jan Kwieci{\'n}ski.

We would like to thank A. Rostworowski for comments and for reading the
notes. M.P. would like to thank W.~Broniowski and E.~Ruiz-Arriola for
comments and discussion. Special thanks are due to K.~Goeke and all members
of Inst. of Theor. Phys. II at Ruhr-University where part of this work was
completed. M.P. acknowledges support of the Polish State Committee for
Scientific Research under grant 2 P03B 043 24.

\end{document}